\documentclass{article}

\usepackage[final]{nips_2016}

\usepackage[utf8]{inputenc} 
\usepackage[T1]{fontenc}    
\usepackage{hyperref}       
\usepackage{url}            
\usepackage{booktabs}       
\usepackage{amsfonts}       
\usepackage{microtype}      
\usepackage{graphicx}

\usepackage{array}
\usepackage{enumitem}
\usepackage{underscore}
\usepackage{lipsum}
\usepackage{verbatim}
\usepackage[fleqn]{amsmath}
\usepackage{amssymb}
\usepackage{mathtools}
\usepackage{listings}
\usepackage{color}
\usepackage{graphicx}
\usepackage{eso-pic}

\title{A generalized Bayesian framework for the analysis of subscription based businesses}

\author{
  Rahul Madhavan \AND Ankit Baraskar \\\\
  Atidiv\\
  Pune, IN-411014 \\
  \texttt{rahul.madhavan@atidiv.com} \\
}

\begin{document}

\maketitle

\begin{abstract}
	We have created a framework for analyzing subscription based businesses in terms of a unified metric which we call SCV (single customer value). The major advance in this paper is to model customer churn as an exponential decay variable, which directly follows from experimental data relating to subscription based businesses. This Bayesian probabilistic model was used to compute an expected value for the revenue contribution of a single user. We obtain an exact closed-form solution for the constant churn model, and an approximate closed-form solution for the exponential decay model. In addition, we define a general methodology for decision making processes using sensitivity analysis of the model equation, which we illustrate with a real-life case study for a food based subscription business.
\end{abstract}

\section{Introduction}
{
	Subscription based businesses (referred to as SBS henceforth) have been around for quite some time - newspapers being one of the very first of these, where a subscription entitles you to the daily news. Post-paid mobile service was one of the early instances of an SBS delivering a service rather than a product on a subscription basis. 
	
	Some of the largest businesses today are exploring the subscription based model - Amazon is attempting to make grocery subscription commonplace. Netflix is fast becoming the alternative television, with Amazon prime video, Hulu and other services vying for a share of this new segment.
	
	One of the reasons for the popularity of an SBS is its guaranteed revenue stream per customer - much like rent in real estate, the company is guaranteed its returns till the customer quits, also referred to as a customer \textit{churning}. Thus, for an existing SBS, there are two primary ways to increase its revenue - customer acquisition and customer retention. In this paper, we model both of these processes and explore how the interplay between them can be optimized for overall revenue by addressing a single metric - the Single Customer Value (SCV).
}

\section{Motivation}
\subsection{Need for a single customer-based metric}
{
	KPI alignment of individual functions towards overall company metrics is a largely un-addressed area for many companies. Individual functions within companies tend to optimize their own goals, often to the detriment of overall company progress. The need for alignment of divisions and functions towards an overall framework drives our idea of a single customer score.
	
	Here, we outline an equation that models the business process for an SBS in terms of both internal as well as any external factors known about customers. This in turn creates a framework where we then have a single metric that we are optimizing for. An optimization differential equation within this framework leads us to the most important action points for the process as a whole, and therefore the business.
	
}	

\subsection{Comparisons to existing models}
{
	Most major attempts to attack complete business modeling have come from the finance side. However, these suffer from four main problems. Firstly, financial models apply a revenue and profit based approach versus our process based approach. Secondly, the low detailing into the process parameters in a financial model limits visibility of outcomes on process changes. Thirdly, financial models tend to focus on internal parameters like rate of return for judging a process change whereas the below model allows both internal and external variations.

	Models for CLV (Customer Lifetime Value) have been developed in the context of non-subscription and subscription based companies$^{[1], [2]}$. However, these models are hard to interpret owing their inherently stochastic nature - here we provide a highly accessible and flexible deterministic model in an SBS context.  
}

\section{Assumptions}
We list the assumptions that go into this model.
{\subsection*{General Assumptions}

\begin{enumerate}
	\item The model is for a single product.
	\item Customer differences can be modeled as categorical variables.
	\item The model does not deal with nonlinear interactions between different features.
\end{enumerate}

\subsection*{Process Unit Assumptions}

\begin{enumerate}
	\item The customer is a single, well-defined entity in the process flow.
	\item The model assumes that a user who has not churned generates fixed revenue per unit time.
	\item The model does not explicitly address cross-customer correlation effects.
\end{enumerate}

\subsection*{Process Assumptions}

\begin{enumerate}
	\item The process is self-contained.
	\item The process parameters are constant across process units.
	\item The process parameters are constant across time.
\end{enumerate}

\subsection*{Customer Flow Assumptions}

\begin{enumerate}
	\item The flow is uni-directional and conserves process units.
	\item Each exit point is differentiated by a unit of subscription time. 
	\item Each exit point has a certain ab-initio probability of exit (churn) associated with it. 
\end{enumerate}

\subsection*{Path Assumptions}

\begin{enumerate}
	\item Each path is uniquely defined by a point of entry a the point of exit.
	\item The ab-initio probability associated with an exit is the conditional probability of exit given the process unit reaches the previous point in the flow.
	\item The expected value of the overall process is a weighted sum of the expected values of all possible paths
\end{enumerate}
}
\begin{table}
	\begin{center}
		\begin{tabular}{@{}ll} \toprule
			Parameter & Definition\\ \midrule
			SCV & Single Customer Value\\ 
			$AC_i$ & Acquisition Path through Channel $i$\\
			Tr & Path to Trial Post Acquisition\\
			$C_{Tr}$ & Path to failure of Trial (month 0)\\
			$C_i$ & Path to failure at month $i$, post success of month $i-1$\\ [1ex] 
			$S_{Tr}$ & Path to success of Trial\\
			$S_i$ & Path to success at month $i$, post success of month $i-1$\\
			$P_{i}$ & Overall probability of path $i$\\
			$p_{i}$ & Conditional probability of path $i$ given success at $i-1$ \\
			$P_{Tr}$ & Probability of entering trial after acquisition\\
			$P_{C_{Tr}}$ & Probability of churn during trial\\
			$P_{S_{Tr}}$ & Probability of continuation from Trial\\
			$P_{C_t}$ & Overall probability of churn after $t$ time periods\\
			$P_{S_t}$ & Overall probability of continuing after $t$ time periods\\
			$CAC_j$ & Cost of Acquiring Customer through channel $j$\\
			$CAC_{mean}$ & Average Cost of Acquiring Customer through all channels\\
			$CC$ & Cost of retaining a customer beyond the trial period\\
			$R_{i}$ & Overall contribution to revenue from path $i$\\
			$r_{i}$ & Conditional revenue given path $i$\\
			$r_{pt}$ & Revenue per unit time net of costs\\
			$R_{C_{Tr}}$ & Contribution to overall net revenue given churn at Trial\\
			$R_{C_t}$ & Contribution to overall net Revenue given churn at time $t$\\
			$r_{C_{Tr}}$ & Net Revenue on the path churning out during trial\\
			$r_{C_t}$ & Net Revenue on the path churning at time $t$\\
			$F$ & Factor variable external the process\\ \bottomrule
		\end{tabular}
	\end{center}
	\caption{Variable definitions used in this work}
	\label{table:vardef}
\end{table}

\section{Setup of Equations}
\begin{figure}[h!]
	\includegraphics[width=\linewidth]{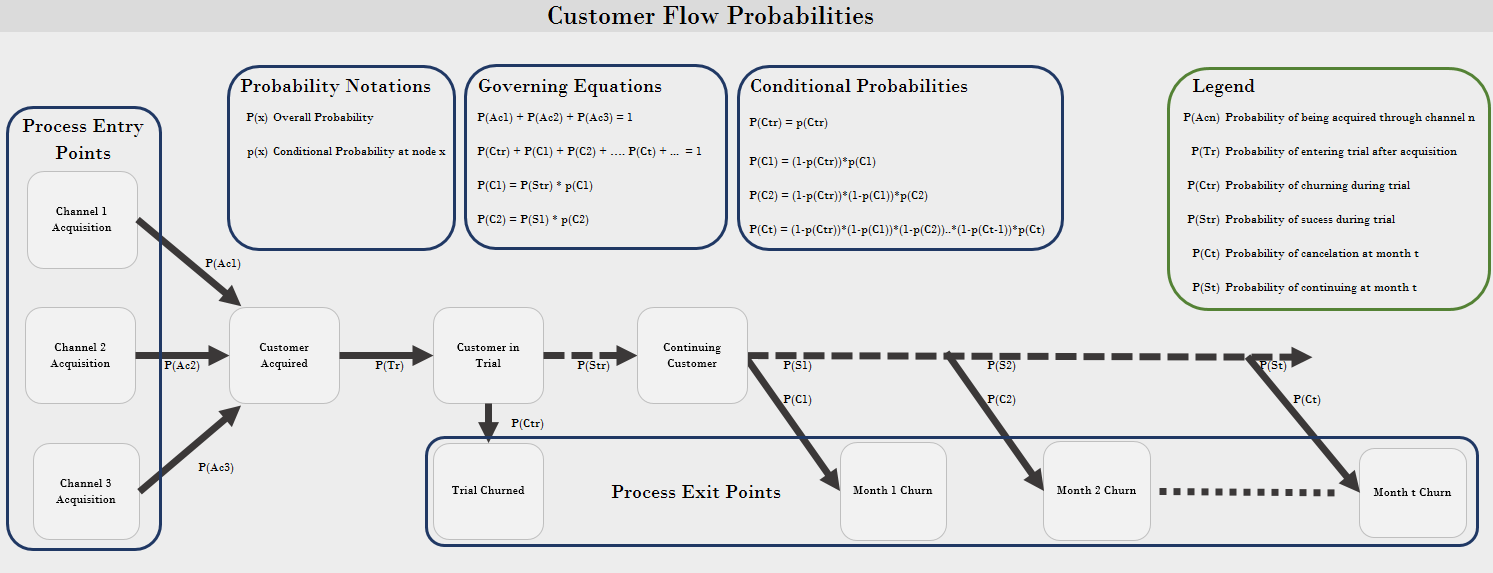}
	\caption{The subscription business as a Bayesian churn model}
	\label{fig:CustomerProcessFlow}
\end{figure}

We set up the equations for our model described in figure \ref{fig:CustomerProcessFlow} using the assumptions defined in the previous section. Refer to Table \ref{table:vardef} for the variable definitions.

\subsection{Probability Equations}

\begin{equation} 
\sum_{i=1}^n P_{AC_i} = 1
\end{equation}
\begin{equation} \label{eq:TotalExitProb}
P_{C_{Tr}} + \sum_{t=1}^\infty P_{C_t} = 1
\end{equation}

\begin{equation} 
P_{S_{Tr}} = 1- p_{C_{Tr}}
\end{equation}
\begin{equation} 
\begin{aligned}
P_{S_i} & = (1-p_{C_{Tr}}) \times \prod_{t=1}^i (1-p_{C_t}) \\  
\end{aligned}
\end{equation} 
\\
Since a customer cannot churn before the end of the trial stage,
\begin{equation} 
P_{C_{Tr}} = p_{C_{Tr}}
\end{equation}
\begin{equation} 
\begin{aligned}
P_{C_t} & = P_{S_{t-1}} \times p_{C_t}
\end{aligned}
\end{equation} 
\begin{equation} 
\begin{aligned}
\implies P_{C_t} & = (1-p_{C_{Tr}}) \times \prod_{i=1}^{t-1}(1-p_{C_i}) \times p_{C_t}
\end{aligned}
\end{equation} 

\subsection{Revenue Equations}
\begin{equation} 
\begin{aligned}
r_{CAC_i} & = P_{AC_i} * CAC_i 
\end{aligned}
\end{equation}
\begin{equation}
\begin{aligned}
CAC_{mean} & = \sum_{i=1}^n r_{CAC_i} \\
& = \sum_{i=1}^n  P_{AC_i} * CAC_i 
\end{aligned}
\end{equation}

\begin{equation}
r_{Tr} = CAC_{mean} + CC\\
\end{equation}
\begin{equation}
\begin{aligned}
R_{Tr} & = P_{C_{Tr}} \times r_{Tr}  \\
& = P_{C_{Tr}} \times (CAC_{mean} + CC)\\
& = p_{C_{Tr}} \times (CAC_{mean} + CC)
\end{aligned}
\end{equation}

\begin{equation}
\begin{aligned}
r_{C_t} & = CAC_{mean} + CC + t \times r_{pt}
\end{aligned}
\end{equation}

\begin{equation}
\begin{aligned}
R_{C_t} & = r_{C_t} \times P_{C_t}
\end{aligned}
\end{equation}
This expands to
\begin{equation} \label{eq:RevValue}
\begin{aligned}
R_{C_t} & = (CAC_{mean} + CC + t \times r_{pt})\times P_{C_t}\\
\end{aligned}
\end{equation}

\begin{equation} \label{eq:SCV}
SCV = \sum_{t=1}^\infty {R_{C_t}} + R_{Tr} \\
\end{equation}

\section{Constant churn probability model}
{
	Here we assume that each point of exit in our process flow has a constant conditional probability of churn post the trial period. This has been found to be a decent approximation for businesses with a good product and high customer retention.
}
\subsection{Probability Equations}
Let
\begin{equation}
p_{C_t} = 1- \theta
\end{equation}
\begin{equation} \label{eq:gammadef}
p_{C_{Tr}} = 1- \gamma
\end{equation}
\noindent Consequently, we can write the overall probabilities using Eq (8) as:
\begin{equation}
\begin{aligned}
P_{C_t} & = (1 - p_{C_{Tr}}) \times \prod_{i = 1}^{t-1} (1 - p_{C_i}) \times p_{C_t}\\
& = (1 - \theta)(1 - p_{C_{Tr}})\theta^{t-1}\\
& = \gamma(1 - \theta) \theta^{t-1}  
\end{aligned}
\end{equation}

\subsection{Value Equations}
\begin{equation}\label{eq:constchurntrrev}
R_{Tr} = P_{Tr} \times (CAC_{mean} + CC)
\end{equation}
\begin{equation}\label{eq:constchurnallrev}
R_{C_t}  = P_{C_t}\times(CAC_{mean} + CC + t \times R_{Pt})
\end{equation}

\subsection{Overall Value}
From equations \ref{eq:SCV}, \ref{eq:constchurntrrev} and \ref{eq:constchurnallrev}
\begin{equation}
\begin{aligned}
SCV & = P_{Tr} \times (CAC_{mean} + CC) + \sum_{t=1}^\infty {R_{C_t}}\\
& = (CAC_{mean} + CC) \times (P_{Tr} + \sum_{t=1}^\infty P_{C_t}) + \frac{R_{Pt} \gamma}{(1 - \theta)}\\
\end{aligned}
\end{equation}
Here, we used the fact that $p_{C_t}$ is constant to arrive at an infinite geometric series, which was condensed into the above equation.
From equation \ref{eq:TotalExitProb}
\begin{equation*}
P_{Tr} + \sum_{t=1}^\infty P_{C_t} = 1
\end{equation*}
\begin{equation}
\begin{aligned}
SCV & = (CAC_{mean} + CC) \times (P_{Tr} + \sum_{t=1}^\infty P_{C_t}) + \frac{R_{Pt} \gamma}{(1 - \theta)}\\
& = (CAC_{mean} + CC) +  \frac{R_{Pt} \gamma}{(1 - \theta)}\\
\end{aligned}
\end{equation}
Re-substituting for $\theta$, we obtain the final SCV equation for the constant churn model as follows:
\begin{equation} \label{eq:finaltcvflatchurnmodel}
SCV = (CAC_{mean} + CC) +  \frac{R_{Pt} \times \  (1-p_{C_{Tr}})}{p_{C_t}}\\
\end{equation}

\section{Exponential decay churn probability model}
\begin{figure}[h!]
	\includegraphics[width=\linewidth]{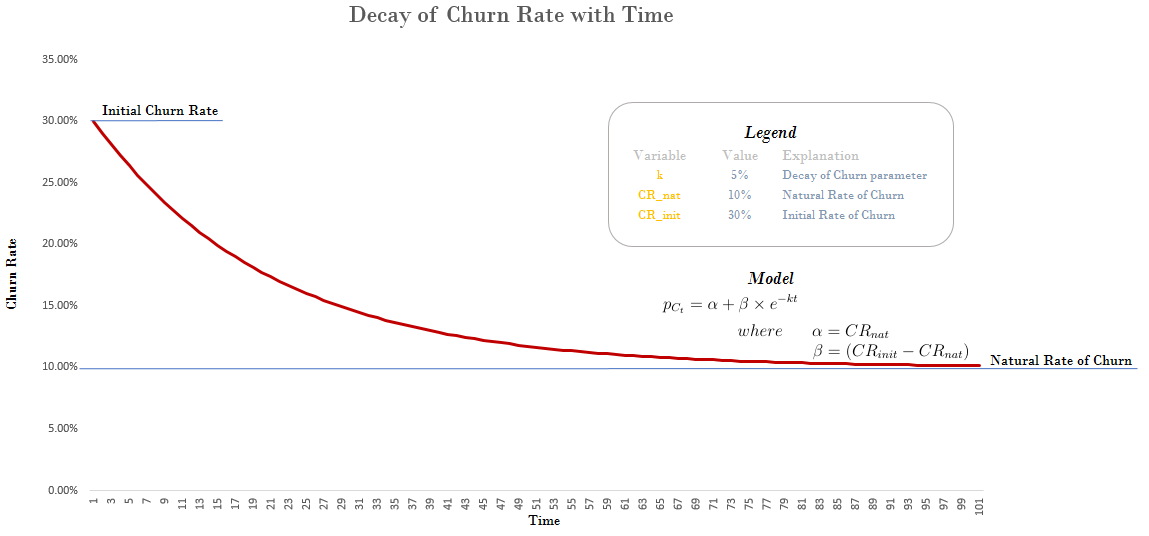}
	\caption{Churn is modelled through an exponential decay from the initial rate to the natural rate}
	\label{fig:exponentialmodel}
\end{figure}

Although the constant churn model is a fairly good approximation for cohort analysis post trial, we observed from experimental data such as that depicted in Figure \ref*{fig:exponentialmodel} that the overall churn rate for an actual SBS is closer to an exponentially decaying quantity. This motivated us to formulate churn in terms of variables defined in Table \ref{table:vardefexp} as follows.

\subsection{Model Formulation}

\begin{table}
	\begin{center}
		\begin{tabular}{@{}ll} \toprule
			Variable & Description \\ \midrule
			$CR_{init}$ & Churn Rate at time period 0 \\
			$CR_{nat}$ &  Natural Rate of Churn to which churn rate decays \\
			$CR_{t}$ & Churn Rate at time period t\\
			$k$ & decay constant describing rate of decay of churn rate\\ \bottomrule
		\end{tabular} 
	\end{center}
	\caption{Additional variable definitions for the exponential model}
	\label{table:vardefexp}
\end{table}

\begin{align}
\alpha = CR_{nat} \\
\beta = (CR_{init} - CR_{nat})\\
CR_{t} = CR_{nat} + (CR_{init} - CR_{nat}) \times e^{-kt}\\
p_{C_t} = \alpha + \beta \times e^{-kt}
\end{align}

\subsection{Extending to Overall Probability Equations}

From Eq \ref{eq:gammadef}, $\gamma  = 1-p_{C_{Tr}}$

\begin{equation}\label{eq:pctbeforereduce}
\begin{aligned}
P_{C_t} & = (1 - p_{C_{Tr}}) \times p_{C_t} \times \prod_{i = 0}^{t-1} (1 - p_{C_i}) \\
& = \gamma  \times (\alpha + \beta \times e^{-k t}) \times \prod_{i = 0}^{t-1} (1 - (\alpha + \beta \times e^{-i\times k}))
\end{aligned}
\end{equation}

\subsection{Analysis of churn rate approximations}
The exponential decay model in its original state proves to be analytically intractable. Consequently, we considered 6 different simplifying assumptions, where we compared the numerical values for the exact solution series and the approximate closed-form solutions. This helped us determine ranges for process parameters wherein these approximations would hold good. Based on the analysis, we then chose the following single simplifying assumption that holds true within reasonable error for an adequate fraction of the valid parameter space.

\begin{equation}\label{eq:alphabetasimplifier}
\begin{aligned}
\gamma \times \prod_{i = 0}^{t-1} (1 - (\alpha + \beta \times e^{-i\times k})) \times (\alpha + \beta \times e^{-k t}) = 
\gamma \times \prod_{i = 0}^{t-1} (1 - (\alpha + \beta )) \times (\alpha + \beta \times e^{-k t})\\
\end{aligned}
\end{equation} 
where
	$\alpha \in (0.001,1),$ 
	$\beta \in (0.001, 1 - \alpha),$
	$\gamma \in (0.001,1),$
	$k \in (0.001,1)$

The following measures were used for validation:
\begin{itemize}
	\item Average: $\overline{\sum_{t=1}^{1000}|P_{C_t}^{model} - P_{C_t}^{truth}|} = 0.0488 $
	\item Min: $\min{(\sum_{t=1}^{1000}|P_{C_t}^{model} - P_{C_t}^{truth}|)} = 1.19*10^{-7} $
	\item Max: $\max{(\sum_{t=1}^{1000}|P_{C_t}^{model} - P_{C_t}^{truth}|)} = 0.5849 $
	\item Stdev: $\sigma{(\sum_{t=1}^{1000}|P_{C_t}^{model} - P_{C_t}^{truth}|)} = 0.0765$
\end{itemize}
 A further analysis of input parameters that gave the highest deviation for this suggests that the maximum absolute deviation for any 1 time period is less than 0.01\%. Large deviations are observed when $k$ is large and $\alpha$, $\beta$ and $\gamma$ are all small. Hence we recommend this model for all parameter values where $k < 1$.
 
\subsection{Overall Probability}
Let
\begin{equation}
1- \alpha = \alpha'
\end{equation}

From Eq: \ref{eq:pctbeforereduce} and Eq: \ref{eq:alphabetasimplifier}, we get
\begin{equation}
\begin{aligned}\label{eq:overprobeqn}
P_{C_t} & = \gamma \times (1-\alpha-\beta)^{t-1} \times (\alpha + \beta \times e^{-k t}) \\
& = \gamma \times (\alpha' - \beta)^{t-1} \times (\alpha + \beta \times e^{-k t}) \\
& = \alpha\ \gamma\ (\alpha' - \beta)^{t-1} + \beta\ \gamma\ (\alpha' - \beta)^{t-1}\ e^{-k t}
\end{aligned}
\end{equation}

\subsection{Overall Value}
Since the underlying process is the same, we can write the $SCV$ expression using equation \ref{eq:SCV} as
\begin{equation}
\begin{aligned}
SCV & = R_{Tr} + \sum_{t=1}^\infty {R_{C_t}}\\
& = P_{Tr} \times (CAC_{mean} + CC) + \sum_{t=1}^\infty {R_{C_t}}\\
& = (CAC_{mean} + CC) \times (P_{Tr} + \sum_{t=1}^\infty P_{C_t}) + R_{Pt} \times \sum_{t=1}^\infty { P_{C_t}\times t}\\
\end{aligned}
\end{equation}
Substituting from equation \ref{eq:TotalExitProb},
\begin{equation}
\begin{aligned}
SCV & = (CAC_{mean} + CC) + R_{Pt} \times \sum_{t=1}^\infty { P_{C_t}\times t}\\
\end{aligned}
\end{equation}
From equation \ref{eq:overprobeqn}: 
\begin{equation*}
P_{C_t} = \alpha\ \gamma\ (\alpha' - \beta)^{t-1} + \beta\ \gamma\ (\alpha' - \beta)^{t-1}\ e^{-k t}
\end{equation*}
Therefore,
\begin{equation}
\begin{aligned}
SCV & = (CAC_{mean} + CC) + R_{Pt} \times \sum_{t=1}^\infty { (\alpha\ \gamma\ (\alpha' - \beta)^{t-1} + \beta\ \gamma\ (\alpha' - \beta)^{t-1}\ e^{-k t}) \times t}\\
& = (CAC_{mean} + CC) + \gamma\ R_{Pt}\  \bigg\{ \frac{\alpha}{{(\alpha + \beta)}^2} + \frac{\beta\  e^{-k}}{((1-\ e^{-k}) + (\alpha+\beta) \ e^{-k})^2}\bigg\} \\
\end{aligned}
\end{equation}
Substituting back $CR_{nat}$ and $CR_{init}$ we get
\begin{equation}\label{eq:finaltcveqn}
\begin{aligned}
SCV & = (CAC_{mean} + CC) + \gamma\ R_{Pt}\  \bigg\{ \frac{CR_{nat}}{{(CR_{init})}^2} + \frac{(CR_{init}-CR_{nat})\  e^{-k}}{((1-\ e^{-k}) + (CR_{init}) \ e^{-k})^2}\bigg\} \\
\end{aligned}
\end{equation}

\section{Partial Differentials of the SCV Equation}
\subsection{Overview}

In the first part of our paper, we have arrived at a single equation describing the entire process as one key metric - the Single Customer Value. We now proceed to determine the sensitivity of the SCV to various factors using equation \ref{eq:finaltcveqn}. Knowing the sensitivity of the model to its various parameters, we can further compute its response to changes in external factor variables by studying their impact on model parameters, thus giving us a powerful decision making tool that does not rely on vague human intuition but on simple numerical computations.

\subsection{Partial Differentials}
From Eq: \ref{eq:finaltcveqn}
\begin{align*}
SCV & = (CAC_{mean} + CC) + \gamma\ R_{Pt}\  \bigg\{ \frac{CR_{nat}}{{(CR_{init})}^2} + \frac{(CR_{init}-CR_{nat})\  e^{-k}}{((1-\ e^{-k}) + (CR_{init}) \ e^{-k})^2}\bigg\}\\\\
& = f(CAC_{mean}, CC, \gamma, R_{pt}, CR_{nat}, CR_{init}, k)\\
\end{align*}
Writing the vector of variables as $\textbf{x} = (CAC_{mean}, CC, \gamma, R_{pt}, CR_{nat}, CR_{init}, k)$, we can compute the sensitivity of $SCV$ to features $F$ using the chain rule:\\
\begin{equation}\label{eq:partderiv}
\begin{aligned}
\frac{\partial SCV}{\partial F} = \sum_{i=1}^7 \frac{\partial SCV}{\partial x_i} \times \frac{\partial x_i}{\partial F}
\end{aligned}
\end{equation}

From Eq: \ref{eq:finaltcveqn}, we can show the partial differentials with respect to each of the sub-variables to the SCV equation.

\begin{equation}\label{eq:partderivcac}
\frac{\partial SCV}{\partial CAC_{mean}} =  1
\end{equation}

\begin{equation}\label{eq:partderivcoc}
\frac{\partial SCV}{\partial CC} =  1
\end{equation}

\begin{equation}\label{eq:partderivptr}
\frac{\partial SCV}{\partial P_{Tr}}=  - R_{Pt} \bigg( \frac{CR_{nat}}{{(CR_{init})}^2} + \frac{(CR_{init}-CR_{nat})\  e^{-k}}{((1-\ e^{-k}) + (CR_{init}) \ e^{-k})^2}\bigg)
\end{equation}

\begin{equation}\label{eq:partderivrpt}
\frac{\partial SCV}{\partial R_{Pt}} = \gamma \bigg( \frac{CR_{nat}}{{(CR_{init})}^2} + \frac{(CR_{init}-CR_{nat})\  e^{-k}}{((1-\ e^{-k}) + (CR_{init}) \ e^{-k})^2}\bigg)
\end{equation}

\begin{equation}\label{eq:partderivcrnat}
\frac{\partial SCV}{\partial CR_{nat}} = \quad \frac{\gamma \ R_{Pt}}{{(CR_{init})}^2} + \frac{-\gamma  R_{Pt} e^{-k}}{((1-\ e^{-k}) + (CR_{init}) \ e^{-k})^2}
\end{equation}

\begin{equation}\label{eq:partderivcrinit}
\frac{\partial SCV}{\partial CR_{init}} =   \gamma \ R_{Pt} \Big(- \frac{2  CR_{nat}}{{(CR_{init})}^3} + \frac{e^{-k}(1-\ e^{-k}) + (2CR_{nat} -\ CR_{init}) \ e^{-2k}}{((1-\ e^{-k}) + (CR_{init}) \ e^{-k})^3}\Big)
\end{equation}

\begin{equation}\label{eq:partderivk}
\frac{\partial SCV}{\partial k} = -\gamma \ R_{Pt} \ ((CR_{init}-CR_{nat})\  e^{-k})\bigg( \frac{(1+(1-CR_{init})e^{-k})}{((1-\ e^{-k}) + (CR_{init}) \ e^{-k})^3}\bigg)
\end{equation}

\subsection{Computing the mean time to churn ($\tau_{mean}$)}
{
The mean time for a customer to be churned can be computed by a simple weighted time average, and can be expressed in terms of known parameters using equations (3) and (37), the latter being where we compute the quantity 

\begin{equation*}
\sum_{t=1}^\infty t P_{C_t} = \gamma \Big\{ \frac{CR_{nat}}{CR_{init}^2} + \frac{(CR_{init} - CR_{nat})e^{-k}}{(1-e^{-k} + CR_{init}e^{-k})^2} \Big\}
\end{equation*}

The mean time to churn for a user can be written as
\begin{equation}
\sum_{t=1}^\infty (t+1) P_{C_t} +1 \times P_{C_Tr}= \sum_{t=1}^\infty t P_{C_t} + \sum_{t=1}^\infty P_{C_t} + P_{C_{Tr}}
\end{equation}
This implies
\begin{equation}\label{eq:taumean}
\tau_{mean} = 1+ \gamma \Big\{ \frac{CR_{nat}}{CR_{init}^2} + \frac{(CR_{init} - CR_{nat})e^{-k}}{(1-e^{-k} + CR_{init}e^{-k})^2} \Big\}
\end{equation}
}

\section{Case Study : Food Based Subscription Business}
\subsection{Overview}
{
The company in question is a small-sized silicon valley based subscription food company offering an externally sourced single food type operating on a monthly subscription basis. They have three kinds of subscription offers, viz. 1  6 and 12 month subscriptions. Most of their uptake is on a monthly payment basis. They also offer prepaid options which customers seldom opt for (\textless 1\% usage). Every month customers get a shipment containing three packets of similar food items. The customer gets an option to choose, but sometimes the company may send over the shipment based on a pre-decided menu.

The company only acquires customers through online channels. Most of their acquisition spend is on channels like Facebook and Google based advertising. Their customers are heavily discounted during the first month, which we refer to as a trial period where the company may end up losing or making a small amount of money net of COGS and shipping costs. Further we note that their initial churn rates are very high, though their longer term customers are much less likely to churn than a new customer into the system. While many of their customers initially opt for a 6m or 12m subscription period, we find that these too cancel their subscription quite often or simply default on their payment.

On this note, we can proceed to analyse the company through the SCV equation lens.
}
\subsection{Churn Timeline}
{
	An essential part of the pipeline to estimating the variables is to plot the percentage active customers by time period. This enables us to estimate the variables. 
	The nature of the curve found, which we expect to be true for most subscription based businesses, was also part of the motivation for the exponential decay churn model

	\begin{figure}[h!]
	\centering
	\includegraphics{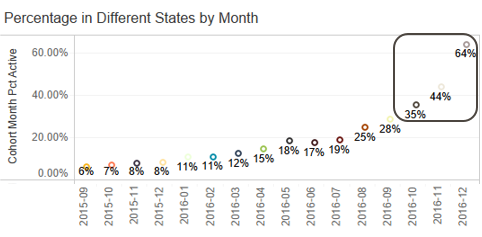}
	\caption{Cohortwise decay of percentage active customers}
	\label{fig:expdata}
	\end{figure}
}
\subsection{SCV computation and local sensitivity analysis}
{
	From equation \ref{eq:finaltcveqn}, we can compute the $SCV$ as -\$31.48. This indicates that the company is losing money, and that the revenue generated per user is much lesser than the cost spent on acquiring them. This leads us to believe that the company should focus on reducing their $CAC$ and improving customer retention. Although it is trying to address the former, the latter issue in our opinion is equally important and can potentially have a much greater impact on the company.

	The values for our model variables were perturbed in order to study the effects on SCV. We computed the gradients as described by our approximate exponential model. We see that the SCV value is most sensitive to the initial and natural (or final) churn rates. Even a 1 \% increase in the initial monthly churn rate can potentially push the SCV down by \$ 1.11, whereas a similar change in the natural churn rate bumps it up by \$ 1.67. Thus, we expect the companies performance to improve significantly if they focus more on customer retention. 

\begin{table}
	\begin{center}
		\begin{tabular}{@{}llr@{}} \toprule
			Variable & Value & Sensitivity\\ \midrule
			$CAC_{mean}$ & -\$35 & 1\\
			$CC$ & $\approx$ \$0 & 1\\
			$R_{P_t}$ & \$6 & 1.58\\
			$P_{Tr}$ & 0.36 & -14.78\\
			$CR_{nat}$ & 0.05 & 167.17 \\
			$CR_{init}$ & 0.30 & -111.81\\
			$k$ & 0.6 & -2.01\\ \bottomrule
		\end{tabular}
	\end{center}
	\caption{Sensitivities computed across different model parameters}
	\label{table:companydata}
\end{table}
}
\subsection{Checking for variability with single feature}
{
	As noted previously, we can use the SCV equation to determine the impact of an external variable on the customer value. As an illustration, we choose a categorical variable that takes three distinct values. This variable does not pertain to any particular path in the process and therefore can be considered an external variable. We show the effect of this variable on the mean months to cancel in the process. This leads us to the difference in SCV for every differential increase in the percentage of customers in one particular bucket. We split this categorical variable into multiple features, each representing one particular category.

	\begin{figure}[h!]
		\centering \includegraphics[width=\linewidth]{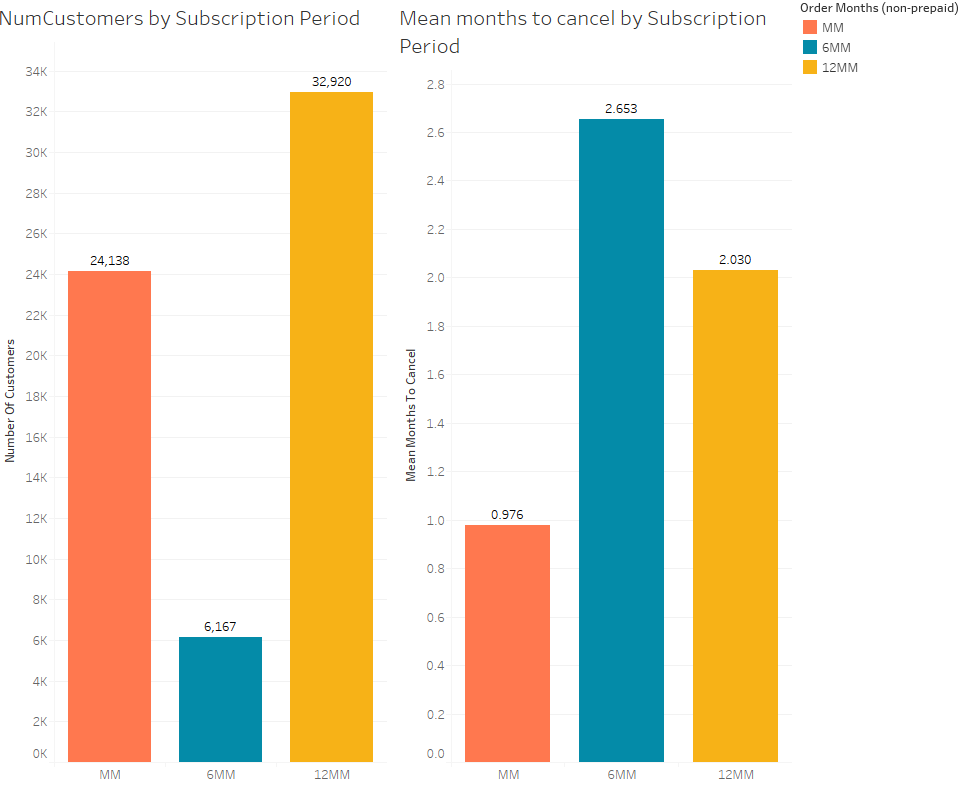}
		\caption{Change in Mean months to cancel by subscription period}
		\label{fig:meanmonthstocancel}
	\end{figure}

	We choose the categorical variable subscription period, and within that, focus our calculation on the 6MM feature bucket and check the effects of having customers move into that bucket. As seen in Figure \ref{fig:meanmonthstocancel}, the mean months to cancel varies across these features. We note that this is the same as the previously defined $\tau_{mean}$ for this process, which for the overall process can be calculated using equation \ref{eq:taumean}, which gives us a value of 1.6884 ($\approx 2$) months. We assume that the natural churn rate and the exponential decay coefficient ($CR_{nat}$ and $k$) are unaffected by a change in the categorical variable. Thus, we can relate change in $\tau_{mean}$ across buckets to change in $CR_{init}$, computed explicitly in the following section.
}
\subsection{Checking for change in SCV by changing Subscription Period}

\begin{equation}
\begin{aligned}
\Delta(SCV) & = \frac{\partial SCV}{\partial CR_{init}} \times \frac{\partial CR_{init}}{\partial F} \times \Delta(F)\\
\end{aligned}
\end{equation}
From Equation \ref{eq:partderivcrinit},
\begin{equation*}
\begin{aligned}
\frac{\partial SCV}{\partial CR_{init}}  = \gamma \ R_{Pt} \Big(- \frac{2  CR_{nat}}{{(CR_{init})}^3} + \frac{e^{-k}(1-\ e^{-k}) + (2CR_{nat} -\ CR_{init}) \ e^{-2k}}{((1-\ e^{-k}) + (CR_{init}) \ e^{-k})^3}\Big)\\
\end{aligned}
\end{equation*}
We note that $\tau_{mean}$ changes from 1.6884 to 2.653. This changes the $CR_{init}$ values from 30\% to 14.83\%\\
In other words
\begin{equation*}
\begin{aligned}
\lim_{\tau\to1.6884} CR_{init} & = 0.3\\
\lim_{\tau\to2.653} CR_{init} & = 0.1483\\
\end{aligned}
\end{equation*}
\begin{equation*}
\begin{aligned}
\Delta(CR) & = \frac{\partial CR_{init}}{\partial F} \times \Delta(F) \\
\Delta(CR) & = 0.1483 - 0.3 = -0.1517\\
\end{aligned}
\end{equation*}
\begin{equation*}
\implies \frac{\partial CR_{init}}{\partial F} \times \Delta(F)  = -0.1517\\
\end{equation*}
We also note the change in partial differentials through these two limits
\begin{equation*}
\begin{aligned}
\lim_{\tau\to1.6884} \frac{\partial SCV}{\partial CR_{init}}  &= -11.14\\
\lim_{\tau\to2.653} \frac{\partial SCV}{\partial CR_{init}}  &= -111.81\\
\end{aligned}
\end{equation*}
\begin{equation*}
\implies \bigg(\frac{\partial SCV}{\partial CR_{init}}\bigg)_{mean} = -(111.81+11.14)/2 = -61.47\\
\end{equation*}
Substituting the values,

\begin{equation*}
\begin{aligned}
\Delta(SCV) & = \frac{\partial SCV}{\partial CR_{init}} \times \frac{\partial CR_{init}}{\partial F} \times \Delta(F)\\
& = \bigg(\frac{\partial SCV}{\partial CR_{init}}\bigg)_{mean} \times \Delta(CR)
& = -61.47 * -0.1517 \\
& = \$9.325
\end{aligned}
\end{equation*}
This implies a SCV of -\$22, or a total change of 30\% on the SCV through a complete change to this feature. Note that this is a linear extrapolation based on the derivative and does not account for non-linear changes.

\section{Conclusions}
{
The churn rates were expressed using an exponential decay model so as to embed the data in our SCV framework. Using an approximation for certain exponential terms lead to a surprisingly clean expression for SCV as a function of parameters such as $R_{pt}$, $P_{C_{Tr}}$ and most importantly $CR_{init}$ - let us call this the AEM (Approximate Exponential Model) as opposed to EM (the "exact" Exponential Model).

A local sensitivity analysis of SCV as computed by the AEM with respect to the feature $CR_{init}$ indicates a high negative value for sensitivity, which is in agreement with our hypothesis as well as the EM. The AEM also allows us to use real data values for the mean time to churn $\tau_{mean}$ in order to exactly back-compute the initial churn rate $CR_{init}$. This relation allows us to tune our initial churn rate in accordance with the mean time to churn.

It should be noted, however, that the AEM works best within the specified parameter bounds. The AEM also shows a lower local sensitivity to $CR_{nat}$, especially in the partial derivative than what we expect, though it is likely due to related parameters $k$ (the decay rate) and $CR_{init}$ not being varied in a more realistic fashion. Further work would involve improving upon the AEM to ensure it performs equally well across greater parameter ranges. More data from similar businesses would also help to verify the model and its underlying assumptions, as well as allow us to develop better and more generalized churn models with predictive value. We hope to integrate this model with a more general approach that allows us to use this single metric along with the number of customers to compute overall company value, with additional factors such as cash flow discounting taken into account in a follow up paper.
}
%
%

\section*{References}

\small

[1] Romero J.\ ,van der Lans R.\ ,Wierenga B.\ (2013) A partially hidden markov model of customer dynamics for CLV measurement, {\it Journal of Interactive Marketing, vol. 27}

[2] McCarthy D.\ ,Fader P.\ ,Hardie B. (2015)  Valuing Subscription-Based Businesses Using Publicly Disclosed Customer Data {\it SSRN Electronic Journal, vol. 81}, pp. 17-35
\end{document}